\documentclass[a4paper,11pt]{article}
\usepackage{pos}
\usepackage[capitalize]{cleveref}
\usepackage{setspace}
\usepackage{titlesec}
\titlespacing\section{0pt}{10pt plus 4pt minus 2pt}{5pt plus 2pt minus 2pt}
\newcommand{\ahvp}{a_\mu^{\rm hvp}}

\newcommand{\be}{\begin{equation}}
\newcommand{\ee}{\end{equation}}
\newcommand{\bea}{\begin{eqnarray}}
\newcommand{\eea}{\end{eqnarray}}

\newcommand{\stat}{\mathrm{stat}}
\newcommand{\syst}{\mathrm{syst}}

\newcommand{\amu}{a_\mu^{\mathrm{hvp}}}

\newcommand{\awin}{a_\mu^{\mathrm{win}}}
\newcommand{\awiniso}{a_\mu^{\mathrm{win,iso}}}
\newcommand{\awinl}{a_\mu^{\mathrm{win,ud}}}
\newcommand{\awins}{a_\mu^{\mathrm{win,s}}}
\newcommand{\awind}{a_\mu^{\mathrm{win,disc}}}
\newcommand{\awinc}{a_\mu^{\mathrm{win,c}}}
\newcommand{\awinisovec}{a_\mu^{\mathrm{win,I1}}}
\newcommand{\awinisosca}{a_\mu^{\mathrm{win,I0}}}
\newcommand{\awinany}{a_\mu^{\mathrm{win,}f}}

\title{Intermediate window observable for the hadronic vacuum polarization contribution to the muon $g-2$ from O$(a)$ improved Wilson quarks}
\ShortTitle{Intermediate window observable for the HVP contribution to the muon $g-2$}

\author[a]{M.\,C\`e}
\author[b]{A.\,G\'erardin}
\author[c]{G.\,von Hippel}
\author[d,e]{R.\,J.\,Hudspith}
\author*[f,e]{S.\,Kuberski}
\author[c,f]{H.\,B.\,Meyer}
\author[f,g]{K.\,Miura}
\author[d,e]{D.\,Mohler}
\author[c]{K.\,Ottnad}
\author[c]{S.\,Paul}
\author[h]{A.\,Risch}
\author[c,f]{T.\,San Jos\'e}
\author[c,f,i]{H.\,Wittig}

\affiliation[a]{Albert Einstein Center for Fundamental Physics (AEC) and Institut für Theoretische Physik, Universität Bern, Sidlerstrasse 5, 3012 Bern, Switzerland}
\affiliation[b]{Aix-Marseille-Universit\'e, Universit\'e de Toulon, CNRS, CPT,
Marseille, France}
\affiliation[c]{PRISMA$^+$ Cluster of Excellence and Institut f\"ur
Kernphysik, Johannes Gutenberg-Universit\"at Mainz, Germany}
\affiliation[d]{Institut für Kernphysik, Technische Universit\"at Darmstadt,
	Schlossgartenstrasse 2, D-64289 Darmstadt, Germany}
\affiliation[e]{GSI Helmholtz Centre for Heavy Ion Research, Darmstadt,
Germany}
\affiliation[f]{Helmholtz-Institut Mainz, Johannes Gutenberg-Universit\"at
	Mainz, Germany}
\affiliation[g]{KEK Theory Center, High Energy Accelerator Research Organization, 1-1 Oho, Tsukuba, Ibaraki 305-0801, Japan}
\affiliation[h]{John von Neumann-Institut f{\"u}r Computing NIC, Deutsches Elektronen-Synchrotron DESY, Platanenallee 6, 15738 Zeuthen, Germany}
\affiliation[i]{Department of Theoretical Physics, CERN, 1211 Geneva 23, Switzerland}

\emailAdd{simon.kuberski@uni-mainz.de}

\preprintnr{MITP-22-102}

\abstract{Following the publication of the new measurement of the anomalous magnetic moment of the muon, the discrepancy between experiment and the theory prediction from the $g-2$ theory initiative
has increased to $4.2\,\sigma$. Recent lattice QCD calculations predict values for the hadronic vacuum polarization contribution that are larger than the data-driven estimates, bringing the Standard Model prediction closer to the experimental measurement. Euclidean time windows in the time-momentum representation of the hadronic vacuum polarization contribution to the muon $g-2$ can help clarify the discrepancy between the phenomenological and lattice predictions.\\
We present our calculation of the intermediate distance window contribution using $N_\mathrm{f}=2+1$ flavors of O$(a)$ improved Wilson quarks. We employ ensembles at six lattice spacings below $0.1\,$fm and pion masses down to the physical value. We present a detailed study of the continuum limit, using two discretizations of the vector current and two independent sets of improvement coefficients.
Our result at the physical point displays a tension of $3.9\,\sigma$ with a recent evaluation of the intermediate window based on the data-driven method.}

\FullConference{%
  The 39th International Symposium on Lattice Field Theory (Lattice2022),\\
  8-13 August, 2022 \\
  Bonn, Germany 
}

\begin{document}
 \maketitle
\section{Introduction}
Given a long-standing tension between experimental findings and Standard Model expectations, the anomalous magnetic moment of the muon, $a_\mu$, is considered to be an excellent probe for physics beyond the Standard Model at the high precision frontier. The combination of the first results of the Fermilab Muon $g-2$ Experiment \cite{Muong-2:2021ojo} with the final result of the E821 experiment at BNL \cite{Bennett:2006fi} yields a $4.2\,\sigma$ discrepancy with the theoretical estimate in the 2020 White Paper \cite{Aoyama:2020ynm}. The uncertainty of this theory prediction is dominated by the uncertainty of the leading-order hadronic vacuum polarization (HVP) contribution, $\ahvp$. The White Paper average for $\ahvp$ with an error of $0.6\%$ is based on evaluations of a dispersion integral involving hadronic cross section data in Refs.~\cite{Davier:2017zfy,
Keshavarzi:2018mgv, Colangelo:2018mtw, Hoferichter:2019gzf,	Davier:2019can, Keshavarzi:2019abf}. Given the foreseen reduction of the experimental uncertainties by upcoming results, the precision of the theory prediction has to improve accordingly in the near future to scrutinize the discrepancy.

Lattice QCD offers the natural framework for an \textit{ab-initio} computation of hadronic contributions to $a_\mu$ and can therefore provide an independent alternative to the traditional data-driven evaluations.
Until recently, the uncertainty of lattice evaluations was too large to have an impact on global averages of $\ahvp$. Thanks to a number of recent algorithmic and conceptual improvements, the evaluation of $\ahvp$ to sub-percent precision is in reach for a number of groups and a first result with $0.8\%$ precision has been published by the BMW collaboration \cite{Borsanyi:2020mff}. This result is in $2.1\,\sigma$ tension with the White Paper average and reduces the tension with the experimental average to $1.5\,\sigma$. To be able to quote a reliable theory prediction for $a_\mu$, this tension between data-driven and lattice estimates has to be understood. Further precise lattice computations are urgently needed.

Time windows in the time momentum representation (TMR) of $\ahvp$ have been introduced in Ref.~\cite{Blum:2018mom}, where a window at intermediate distance has been identified as being ideally suited for a lattice evaluation. It is therefore a good testing ground to compare different lattice calculations at high precision. Furthermore, the evaluation of the same quantity with data-driven methods helps to shed light on the current discrepancies within theory predictions for $\ahvp$. In these proceedings, we summarize the findings of our work in Ref.~\cite{Ce:2022kxy} and discuss their implications.

\section{Lattice setup}
\begin{figure}[t]
	\centering
	\includegraphics*[width=0.48\linewidth]{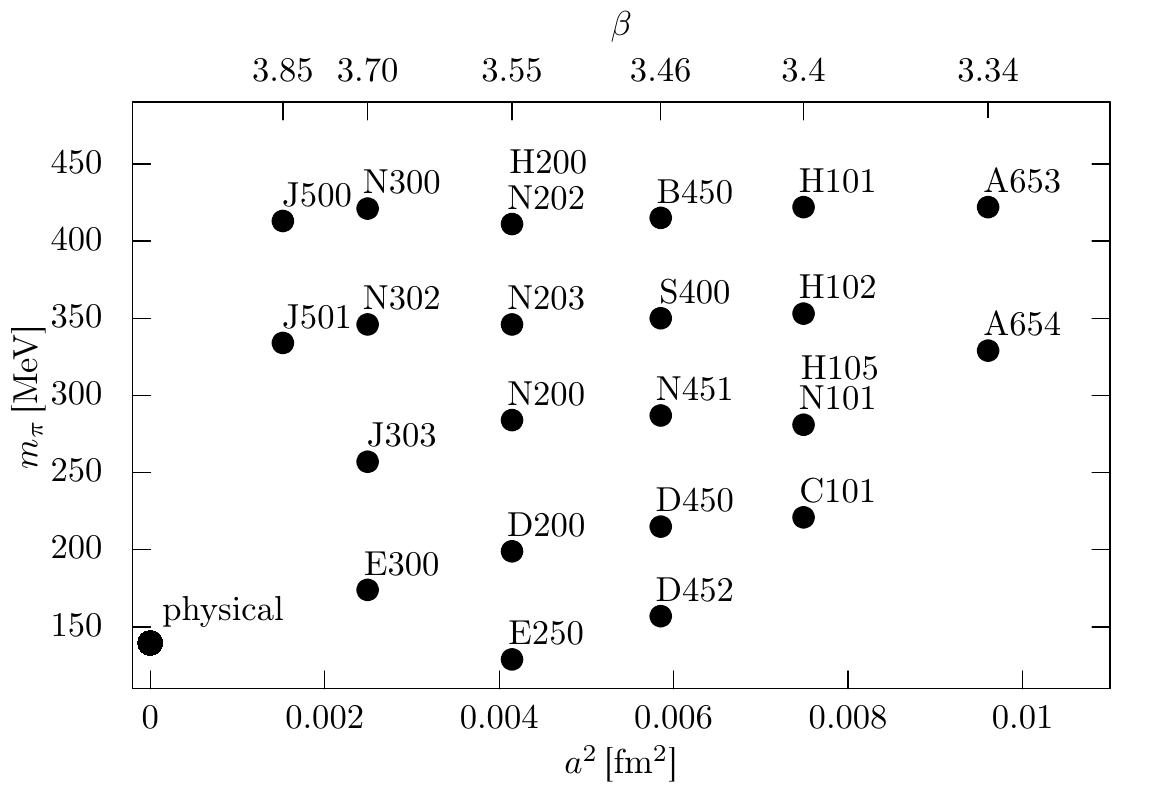}
	\includegraphics*[width=0.48\linewidth]{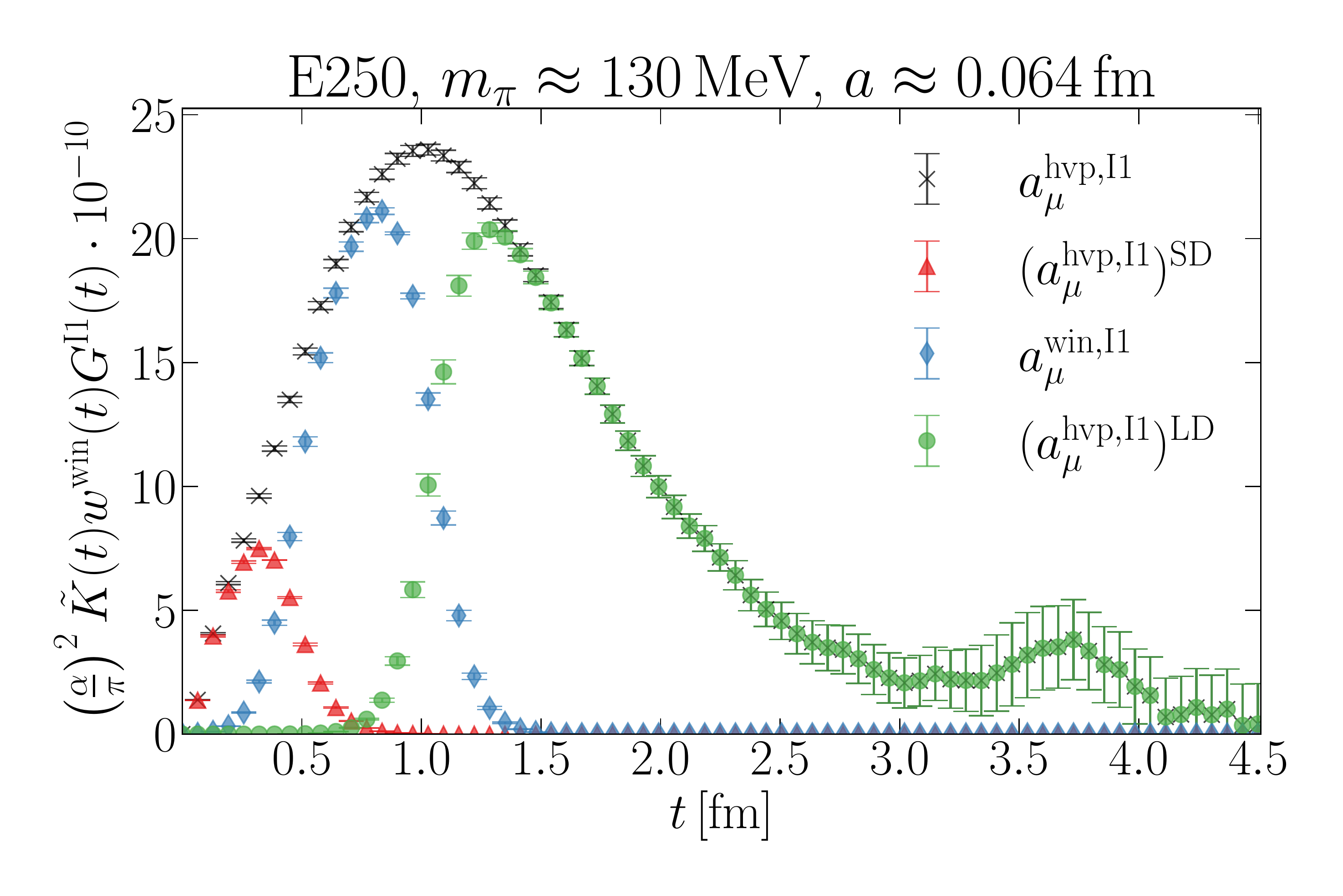}
	\caption{\textit{Left:} Overview of the ensembles used in this work. Two labels for one circle indicate two ensembles with identical parameters but different volumes. \textit{Right:} TMR integrand for the isovector contribution to $\ahvp$ (black crosses) at physical pion mass together with the short (SD), intermediate (win) and long-distance (LD) contributions.}
	\label{fig:setup}
\end{figure}

We work with $2+1$ dynamical flavors of $\mathrm{O}(a)$ improved Wilson fermions and a tree-level improved Lüscher-Weisz gauge action in the isospin limit of QCD on ensembles by the Coordinated Lattice Simulations (CLS) initiative \cite{Bruno:2014jqa}. Our set of 24 ensembles covers six lattice spacings in the range [0.039 - 0.099]\,fm. 
The pion masses are found to be between 130\,MeV and 420\,MeV. On each chiral trajectory, the sum of the bare quark masses is held constant, leading to a constant $\mathrm{O}(a)$ improved bare coupling $\tilde{g}_0$. We employ open boundaries in the temporal direction to alleviate the freezing of the topological charge \cite{Luscher:2011kk}, especially on the finest ensembles. An overview of the ensembles used in this work can be found on the left panel of \cref{fig:setup}.

We compute the intermediate window contribution $\awin$ to $\amu$ in the TMR \cite{Bernecker:2011gh},
\begin{align}
\awin \equiv \left(\frac{\alpha}{\pi}\right)^2\int_0^{\infty} dt \,
\widetilde{K}(t)\,G(t)\,[\Theta(t,t_0,\Delta)-\Theta(t,t_1,\Delta)] \label{def:ID}\,,
\end{align}
from the spatially summed, zero-momentum correlation function $G(t)$ of the electromagnetic current, with a known QED weight function $\tilde{K}(t)$ \cite{DellaMorte:2017dyu} and the smoothed step function $\Theta$ \cite{Blum:2018mom},
\begin{align}
G(t)=-\frac{a^3}{3}\sum_{k=1}^3\sum_{\vec{x}}\left\langle
j_k^{\rm em}(t,\vec{x})\,j_k^{\rm em}(0) \right\rangle, 
\qquad \Theta(t,t^\prime,\Delta) \equiv {\textstyle\frac{1}{2}}\left(
1+\tanh[(t-t^\prime)/\Delta]\right)\,,
\label{eq:TMRcordef}
\end{align}
with $t_0=0.4\,{\rm fm}$, $t_1=1.0\,{\rm fm}$ and $\Delta=0.15\,{\rm fm}$. On the right panel of \cref{fig:setup} we illustrate the integrand of \cref{eq:TMRcordef} (blue diamonds) together with the corresponding integrand for $\ahvp$ (black crosses), as well as for the short- and long-distance contributions to the isovector contribution to $\ahvp$. The noisy long-distance tail of the integrand as well as the short-distance region which is the source of potentially large cutoff effects, are suppressed in $\awin$. Furthermore, the sizable finite-volume effects on $\ahvp$ affect mostly the long-distance tail and are therefore reduced in the case of $\awin$. We find relative statistical uncertainties at the few per-mil level.

We employ two discretizations of the vector current: The local and the point-split version. While only the former needs to be renormalized, both currents have to be $\mathrm{O}(a)$ improved. We utilize two sets of improvement coefficients and renormalization constants, set 1 based on Ref.~\cite{Gerardin:2018kpy} and set 2 based on Refs.~\cite{Heitger:2020zaq,Fritzsch:2018zym}. Both sets remove $\mathrm{O}(a)$ cutoff effects but higher order lattice artifacts differ between the two, providing us with insight in our ability to perform reliable continuum extrapolations. Before extrapolating our results to the continuum limit and interpolating them to physical quark masses, we correct the isovector contribution for finite-size effects. As in Ref.~\cite{Ce:2022eix}, we employ two procedures: At long distances $t > (m_\pi L/4)^2/m_\pi$, where only a few states contribute significantly to the finite-volume isovector correlation function, we compute the difference between finite and infinite-volume correlation function via the Meyer-Lellouch-Lüscher formalism \cite{Meyer:2011um,Luscher:1991cf,Luscher:1990ux} and a Gounaris-Sakurai parametrization \cite{Gounaris:1968mw} of the time-like pion form factor. At short distances that are more relevant for the intermediate window, we employ the method by Hansen and Patella \cite{Hansen:2019rbh,Hansen:2020whp} based on a monopole parametrization of the electromagnetic pion form factor in the space-like region \cite{QCDSFUKQCD:2006gmg}. The resulting finite-size corrections are of the same order as the statistical errors on each ensemble.

We extrapolate the isovector, isoscalar (without charm content) and the charm contribution separately to the physical point according to the following functional form 
\begin{multline}
\awinany(X_a,X_{\pi}, X_K) = \awinany(0,X_{\pi}^{\exp},X_{K}^{\exp}) +
\beta_2 \, X_a^2 + \beta_3 \, X_a^3 + \delta \, X_a^2 X_{\pi}  +
\epsilon \, X_a^2 \log X_a \\ +  \gamma_0 \left( X_K - X_K^{\rm phys}
\right) + \gamma_1 \, \left( X_{\pi} - X_{\pi}^{\exp} \right)  +
\gamma_2 \left( f_{\rm ch}(X_{\pi}) -  f_{\rm ch}(X_{\pi}^{\exp})\right) \,,
\label{eq:fit}
\end{multline}
where \textit{f} denotes the flavor/isospin component and $X_a = a / \sqrt{t_0}$ parametrizes the lattice spacing. The dimensionless variables $X_\pi \propto m_\pi^2$ and $X_K \propto m_K^2 + \frac{1}{2}m_\pi^2$ are employed for the interpolation to physical quark masses, and higher order effects in $X_\pi$ are described via one of the functions $f_{\rm ch}(X_\pi) \in \{0;~ \log(X_\pi);~ X_\pi^2;~ 1/X_\pi;~ X_\pi \log(X_\pi) \}$. We are not able to determine all of the parameters in \cref{eq:fit} in a single fit. Instead, we test variations of the fit form by setting some of the parameters $\beta_3$, $\delta$ and $\epsilon$ to zero, by varying the functional form $f_{\rm ch}$ and by performing cuts in the pion mass and/or the lattice spacing. Our final estimate for the central value, the statistical and the systematic uncertainty of the observable $\awinany$ are determined from a model average \cite{Jay:2020jkz} of the fit results and their respective fit qualities.

\section{Results}
\begin{figure}[t]
	\centering
	\includegraphics*[width=0.48\linewidth]{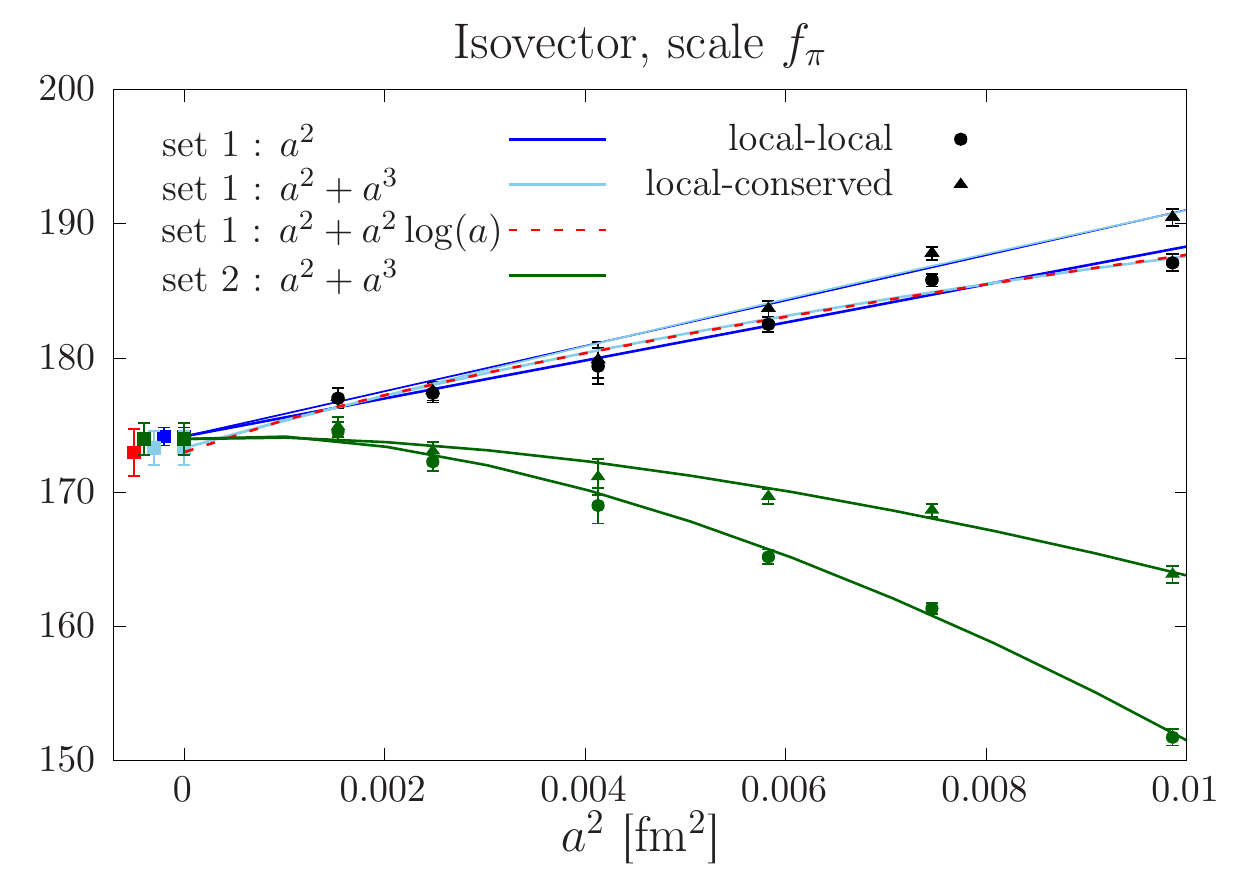}
	\includegraphics*[width=0.46\linewidth]{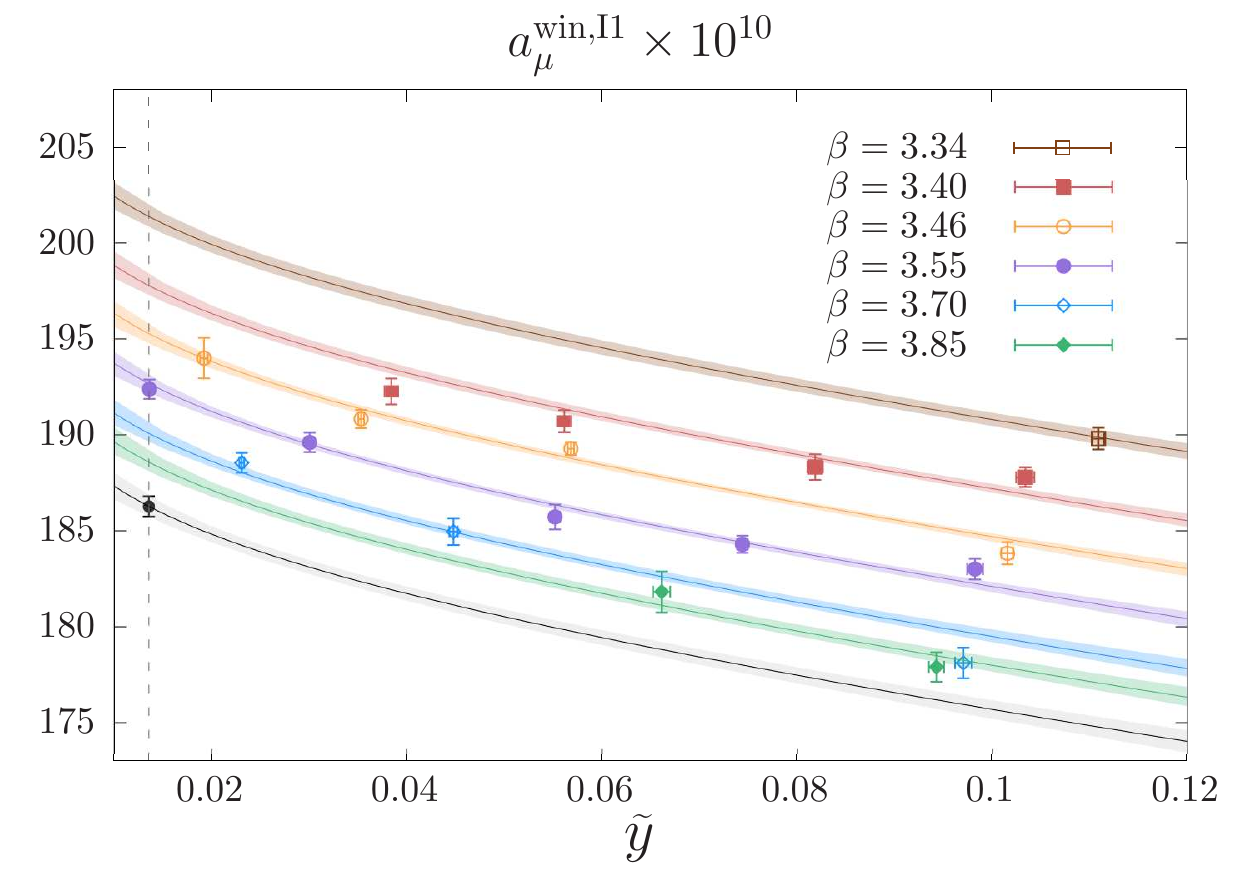}
	\caption{\textit{Left:} Study of the continuum extrapolation of $\awinisovec$ at the $\mathrm{SU}(3)_{\rm f}$ symmetric point. The black and green data points correspond to the two sets of improvement coefficients. \textit{Right:} Exemplary chiral-continuum extrapolation of $\awinisovec$. Each color indicates one value of the bare coupling. The curves show the fit function evaluated at the corresponding lattice spacing. Data are shifted to physical $X_K$. Figures taken from \cite{Ce:2022kxy}.}
	\label{fig:isoevc}
\end{figure}

On the left panel of \cref{fig:isoevc} we illustrate the continuum extrapolation of the dominant isovector contribution to $\awin$ at the $\mathrm{SU}(3)_{\rm f}$ symmetric point, i.e., on the ensembles where $m_\pi = m_K \sim 420\,\mathrm{MeV}$. We show four sets of data based on the two discretization prescriptions of the vector current and the two sets of improvement and renormalization procedures. Whereas the cutoff effects differ substantially between the four data sets, we achieve consistent independent extrapolations to the continuum limit. The universality of the continuum limit therefore provides a strong check of our extrapolations. We note in passing that the data based on set 1 may be extrapolated with a single term $\propto a^2$ over the full range of resolutions. Despite our good control over the continuum limit, the variation of the ansatz for the continuum extrapolation contributes dominantly to the systematic uncertainty of our final result. 

\begin{figure}[t]
	\centering
	\includegraphics*[width=0.48\linewidth]{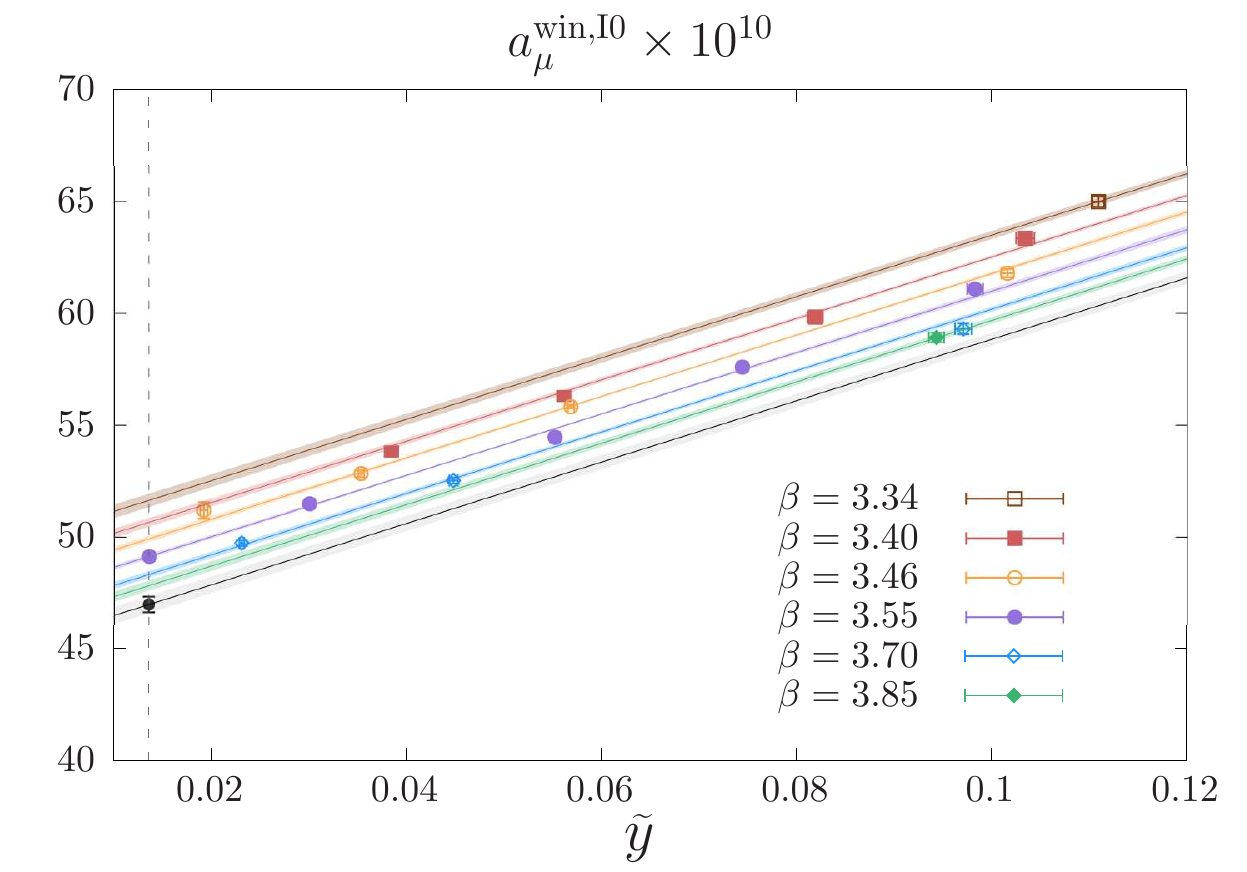}
	\includegraphics*[width=0.46\linewidth]{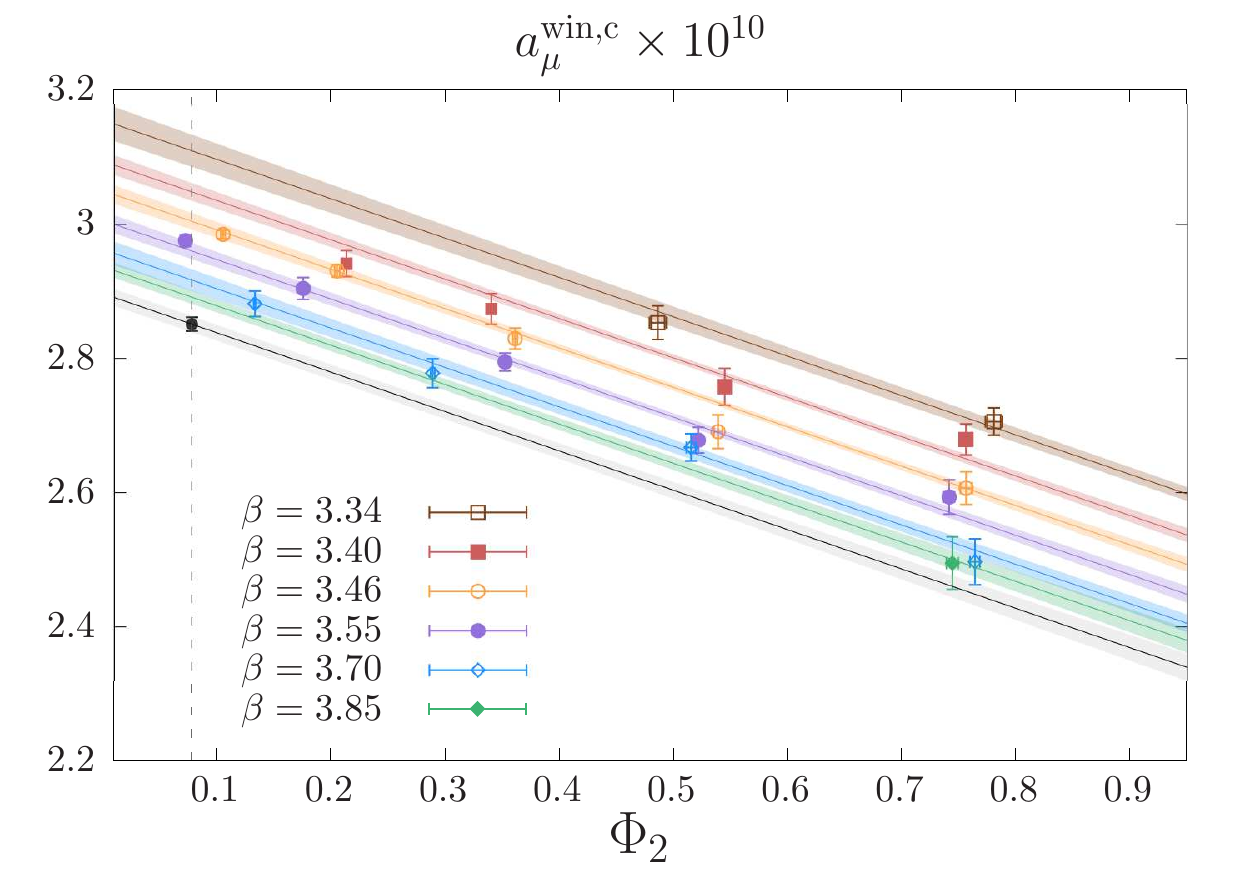}
	\caption{Chiral-continuum extrapolation of contributions to $\awin$. Each color indicates one value of the bare coupling. The curves show the fit function evaluated at the corresponding lattice spacing. Data are shifted to physical $X_K$. \textit{Left:} Isoscalar contribution. \textit{Right:} Charm-connected contribution extrapolated in $X_\pi = \Phi_2 = 8 t_0 m_{\pi}^2$. Figures taken from \cite{Ce:2022kxy}.}
	\label{fig:isosca}
\end{figure}

On the right panel of \cref{fig:isoevc} we illustrate a typical chiral-continuum fit using $f_{\rm ch} = 1/\tilde{y}$ with $\tilde{y} = m_\pi^2/(8\pi f_\pi^2)$ to our data for $\awinisovec$, where the dependence on $X_K$ has been projected out in the plot. The data is well described over the full range of pion masses and, most importantly, constrained by the ensembles close to physical quark masses. Performing variations in the fit form and excluding data at large pion masses does not lead to a significant variation of the result at the physical point. After taking the model average of our fits, we find
\begin{align}
	\awinisovec &= (186.30 \pm 0.75_{\stat} \pm 1.08_{\syst}) \times 10^{-10}\,.       \label{res:I1}
\end{align}
An example for a chiral-continuum extrapolation of the data for the
isoscalar contribution excluding the charm quark is shown on the left
panel of \cref{fig:isosca}. Although the noisy quark-disconnected contribution enters for ensembles away from the $\mathrm{SU}(3)_{\rm f}$ symmetric point, we obtain precise data thanks to the suppression of long-distance contributions. We restrict the model average to fits based on functions $f_{\rm ch}$ that are not singular in the chiral limit and arrive at
\begin{align}
	\awinisosca{}^{,c\!\!\!/} &= (47.41 \pm 0.23_{\stat} \pm 0.29_{\syst}) \times 10^{-10}\,.       \label{res:I0}
\end{align}

The charm-connected contribution is calculated in the partially quenched setup on our $2+1$ flavor configurations. We compute the vector current at three values of the quark mass close to the charm quark mass and perform an interpolation to the point where the mass of the ground-state ${\rm c\bar{s}}$ pseudoscalar meson matches the physical $D_{\rm s}$ meson mass. 
We employ a massive renormalization scheme.
Due to large cutoff effects in the local-local discretization of the correlation function, we take only the local-conserved one into account in our fits. Since the strange quark mass is not held constant along our chiral trajectory, we have to perform a mild chiral extrapolation of the charm-connected contribution.\footnote{Fixing the charm quark mass via the quark-connected contribution to the $\eta_\mathrm{c}$ meson or via the flavor-averaged combination $m_{\bar{D}} = \frac{2}{3} m_D + \frac{1}{3} m_{D_\mathrm{s}}$, as in Ref.~\cite{Chao:2022xzg}, could significantly reduce the pion mass dependence as both masses are approximately constant on our chiral trajectory where $2 am_\mathrm{l} + am_\mathrm{s}$ is held constant. For both choices, no visible dependence of the charm quark mass on the light quark masses has been found on our chiral trajectory in Ref.~\cite{Heitger:2021apz}.}
After performing the model average, we obtain
\begin{align}
	\awinc = (2.89 \pm 0.03_{\stat} \pm 0.03_{\syst} \pm 0.13_{\rm scale}) \times 10^{-10} \,. \label{res:c}
\end{align}
As detailed in Appendix D of Ref.~\cite{Ce:2022kxy}, we estimate the effect of neglecting charm quarks in the sea to be well below our uncertainties. We furthermore neglect the bottom quark contribution to $\awin$ that is expected to be much smaller than our current uncertainty \cite{Colquhoun:2014ica}.

We work in the isospin-symmetric setup of QCD. In order to compare our computation with Nature at the sub-percent level, the effects of the non-degeneracy of the up- and down-quark masses and QED have to be taken into account. We have performed a computation of $\awin$ in QCD+QED using the
technique of Monte Carlo reweighting~\cite{Ferrenberg:1988yz,Duncan:2004ys,Hasenfratz:2008fg,Finkenrath:2013soa,deDivitiis:2013xla} combined with a leading-order perturbative expansion of QCD+QED around isosymmetric QCD in terms of the electromagnetic coupling $e^{2}$ as well as the shifts in the bare quark masses $\Delta m_u,\Delta m_d,\Delta m_s$~\cite{deDivitiis:2013xla,Risch:2021hty,Risch:2019xio,Risch:2018ozp,Risch:2017xxe}. A detailed description of our setup can be found in Refs.~\cite{Risch:2021nfs,Risch:2021hty,Risch:2019xio}. 
Since the renormalization procedure differs from the one used in the isosymmetric QCD calculation, we compute the relative correction due to isospin breaking in the QCD+QED setup. So far, we have performed our computation on five ensembles at three resolutions and pion masses in the range $215$-$352\,\mathrm{MeV}$. The results are displayed on the left panel of \cref{fig:cmpwinfull}. Without performing an explicit extrapolation to the physical point, we estimate the correction to be $(0.3\pm 0.1)\%$ of the isosymmetric contribution. We currently neglect the effect of quark-disconnected diagrams as well as isospin-breaking effects in sea-quark contributions. Furthermore, an investigation of finite-volume effects on the correction is in progress. We double the uncertainty of our estimate to account for these unknown systematic effects before including the correction in our final result.

Combining the results of \cref{res:I1,res:I0,res:c}, we find 
\begin{align}
\awiniso = \awinisovec + \awinisosca{}^{,c\!\!\!/} + \awinc &= (236.60 \pm 0.79_{\stat} \pm 1.13_{\syst} \pm 0.05_{\rm Q} ) \times 10^{-10}\,,
\end{align}
where an additional uncertainty due to the quenching of the charm quark is included. Our final result, after including our estimate of isospin-breaking corrections, is
\begin{align}
	\awin= (237.30 \pm 0.79_{\stat} \pm 1.13_{\syst} \pm 0.05_{\rm Q} \pm0.47_{\rm IB} ) \times 10^{-10}\,.
	\label{eq:finalfull}
\end{align}

\section{Comparison of lattice results}
\begin{figure}[t]
	\centering
	\includegraphics*[width=0.48\linewidth]{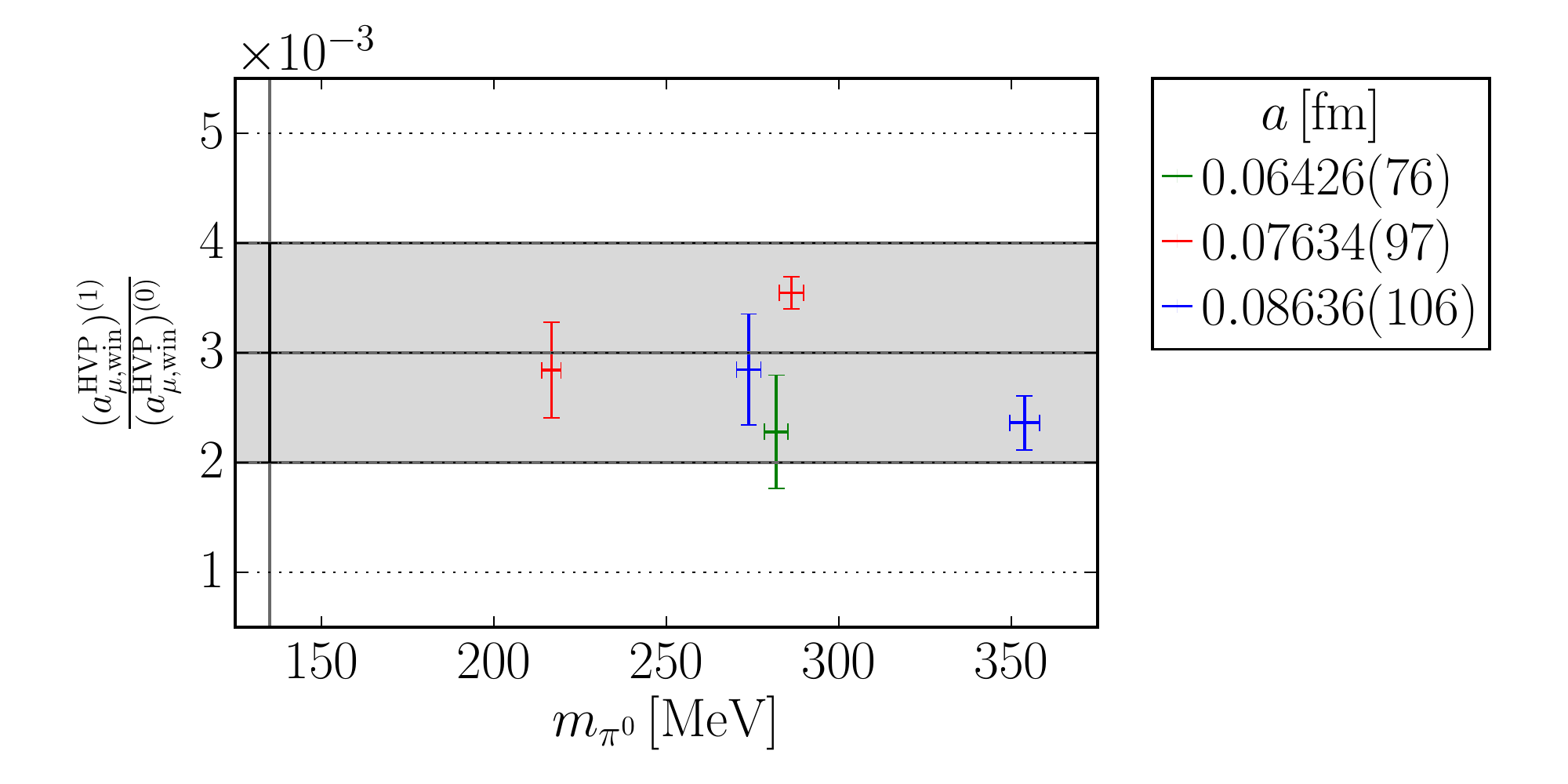}
	\includegraphics*[width=0.43\linewidth]{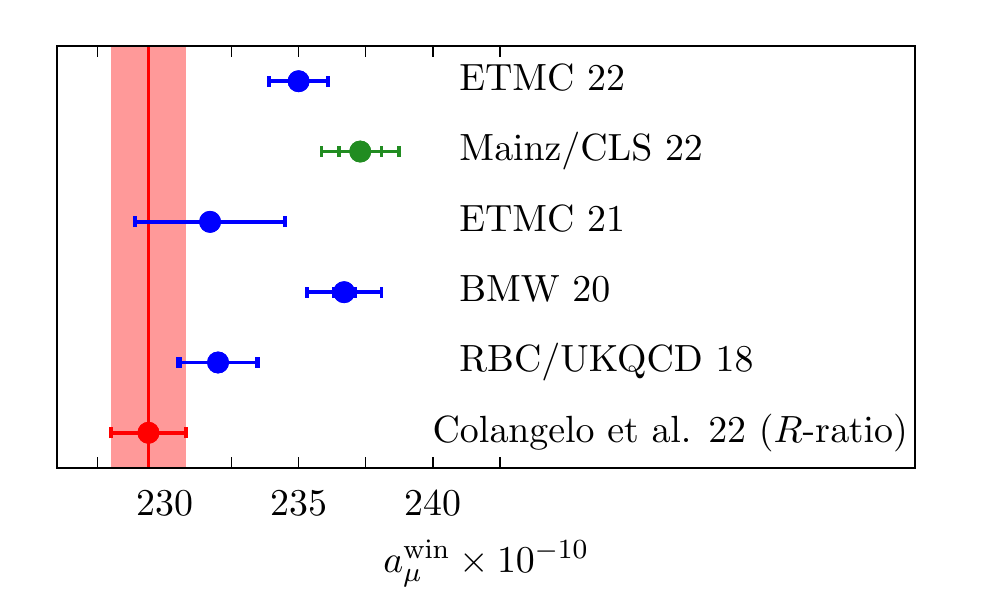}
	\caption{\textit{Left:} Overview of isospin-breaking effects on $\awin$. \textit{Right:} Comparison of our result for $\awin$ including
		isospin-breaking corrections with the estimates by the ETM
		\cite{Giusti:2021dvd,Alexandrou:2022amy}, 
		BMW \cite{Borsanyi:2020mff} and RBC/UKQCD
		\cite{Blum:2018mom} collaborations. 
		The estimate based on the data-driven method of
		Ref.~\cite{Colangelo:2022vok} is shown in red. }
	\label{fig:cmpwinfull}
\end{figure}
\begin{figure}[t]
	\centering
	\includegraphics*[width=0.98\linewidth]{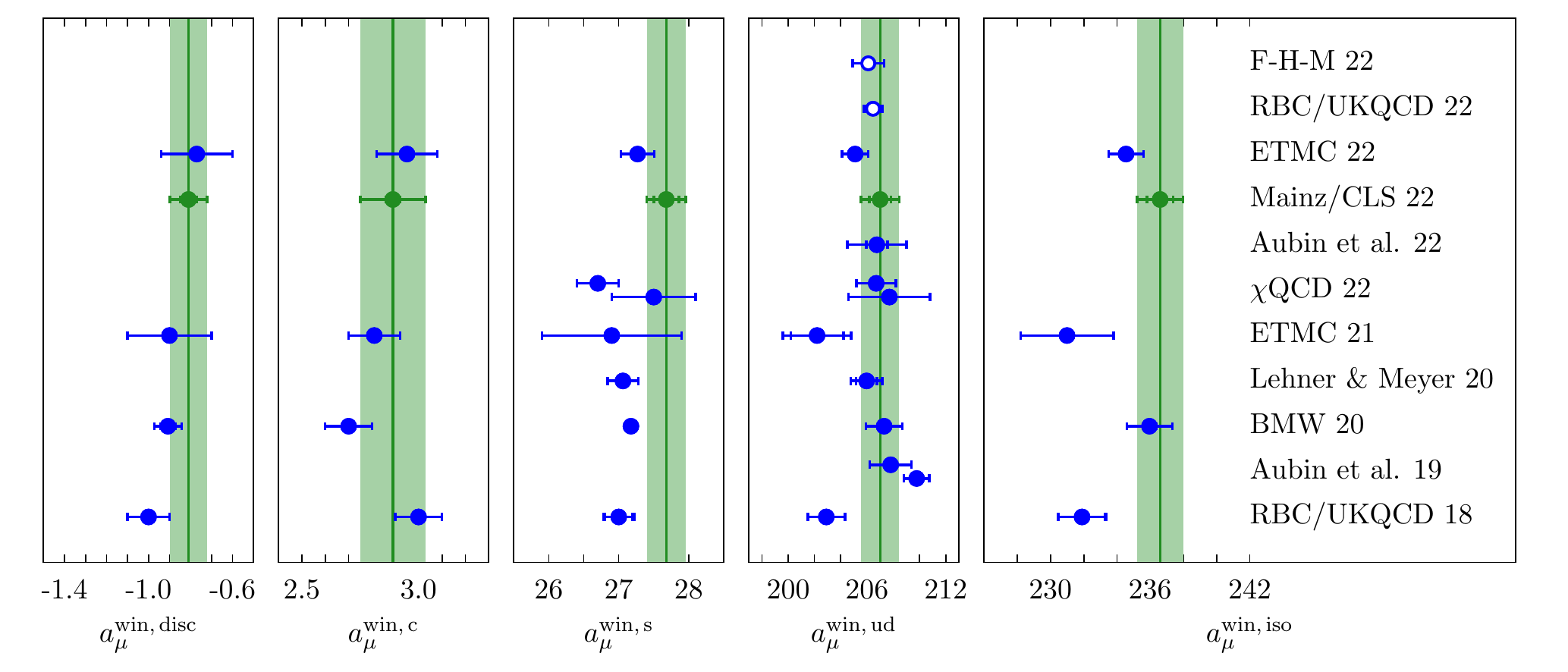}
	\caption{Comparison of our results \cite{Ce:2022kxy} (in units of $10^{-10}$)
		with other lattice calculations \cite{Blum:2018mom, Aubin:2019usy,
			Borsanyi:2020mff, Lehner:2020crt, Giusti:2021dvd, Wang:2022lkq,
			Aubin:2022hgm,Alexandrou:2022amy,Lehner:Edinburgh22,Gottlieb:Benasque22} in isosymmetric QCD. The four panels
		on the left show compilations of the individual quark-disconnected, charm,
		strange and light quark contributions. The result for $\awin$ in
		the isosymmetric case is shown in the rightmost panel. Our results are
		represented by green circles and vertical bands. Results that so far
		have only been presented at workshops are indicated by open symbols.}
	\label{fig:cmpwiniso}
\end{figure}

To compare our results with the findings of other collaborations, we collect them in \cref{fig:cmpwiniso} in the flavor decomposition instead of the isospin decomposition that we have discussed before.\footnote{Note that the sum $\awinisovec + \awinisosca{}^{,c\!\!\!/}$ is very well compatible with $\awinl+\awins+\awind$ in our work, providing an additional cross-check of our chiral-continuum extrapolations.}
Since the writing of Ref.~\cite{Ce:2022kxy}, three additional sets of results have appeared. The calculation in Ref.~\cite{Alexandrou:2022amy} provides results for all flavor components for the intermediate and the short-distance windows. The results of Refs.~\cite{Lehner:Edinburgh22,Gottlieb:Benasque22} for the light-connected contribution have so far only been presented at workshops. This light-connected contribution dominates $\awin$, contributing about $87\%$ to the total.

Let us first consider the subleading contributions $\awind$, $\awinc$ and $\awins$. Here, the results of the different collaborations broadly agree, apart from slight tensions in $\awins$. These tensions are not large enough to have a significant impact on $\awin$. 
The results in Refs.~\cite{Alexandrou:2022amy,Lehner:Edinburgh22} shift the discussion concerning the status of $\awinl$ considerably.
The results labeled RBC/UKQCD 18 \cite{Blum:2018mom} and ETMC 21 \cite{Giusti:2021dvd} based on domain wall fermions and Wilson twisted-mass fermions, respectively, deviate from the bulk of the results for $\awinl$. In both cases, the extrapolation to the continuum limit is quite long and based on a small number of lattice spacings. In ETMC 21, ensembles with pion masses larger than 220\,MeV have been used to compute $\awin$. The new result ETMC 22 employs three ensembles around the physical pion mass and therefore, no chiral extrapolation is necessary. With respect to RBC/UKQCD 18, data at a third, finer lattice spacing (about 0.073\,fm) at physical pion mass as well as a second discretization of the vector current has been added to the analysis in RBC/UKQCD 22. If one takes into account these two updates, the agreement for $\awinl$ between the different groups, working with a wide variety of fermion actions and strategies to approach the physical point, is excellent.\footnote{We note that the comparison presented here contains an inherent ambiguity regarding the definition of the physical point in isosymmetric QCD, see the contributions \cite{Portelli:2022lattice,Tantalo:2022lattice} to this conference.}

Based on the current status displayed in \cref{fig:cmpwiniso}, there is little room for a significant shift in the value for $\awin$ from lattice QCD. This is particularly important when our result, corrected for quark-connected isospin-breaking and electromagnetic effects, and the result of Ref.~\cite{Borsanyi:2020mff} are compared with a recent data-driven evaluation of the same quantity, see the right panel of \cref{fig:cmpwinfull}. A significant tension, $3.9\,\sigma$ between our result and the result of Ref.~\cite{Colangelo:2022vok}, is found.
The absolute deviation between our result for $\awin$ and the prediction in Ref.~\cite{Colangelo:2022vok} is about half of the size of the deviation between the White Paper average for $\ahvp$ and the lattice evaluation of the BMW collaboration. Before a solid statement regarding the Standard Model prediction for $\ahvp$ can be made, this discrepancy between data-driven and lattice evaluations has to be understood.

\section{Outlook}
The foreseen reduction of the experimental uncertainties for $a_\mu$ requires a corresponding improvement of the precision of the SM prediction for $\ahvp$. We aim to contribute to this task by providing a determination of $\ahvp$ to sub-percent precision in the near future. The first precise lattice result in Ref.~\cite{Borsanyi:2020mff} has opened up new questions due to a significant deviation from the well-established dispersive evaluations. As a consequence, time windows in the Euclidean time integral of the TMR are considered to be an ideal testbed to scrutinize the validity of lattice results. For the intermediate-distance window, the cross-check of lattice results has been very successful. However, the comparison with a data-driven evaluation of this quantity points to an even more significant tension than in the case of $\ahvp$.
A similar deviation has been found for the closely related hadronic running of the electromagnetic coupling in Ref.~\cite{Ce:2022eix}.

The investigation of other time windows than the one considered in this work may help to shed light on the origin of the aforementioned discrepancies, see also the recent suggestions in Refs.~\cite{Colangelo:2022vok,Boito:2022njs}. The computation of the short-distance contribution to $\ahvp$ may help to probe the continuum extrapolation of lattice results that makes up a significant fraction of the systematic uncertainty of recent studies. To reach our goal of a sub-percent precision calculation of $\amu$, the main task is to decrease the statistical uncertainty of our calculation, especially at close-to-physical quark masses. Noise reduction techniques in the computation of the vector correlation function, as well as dedicated spectroscopy studies \cite{Andersen:2018mau,Gerardin:2019rua,Paul:2021pjz} will help us to achieve this goal.

\acknowledgments
{
Calculations for this project have been performed on the HPC clusters
Clover and HIMster-II at Helmholtz Institute Mainz and Mogon-II at
Johannes Gutenberg-Universität (JGU) Mainz, on the HPC systems
JUQUEEN, JUWELS and JUWELS Booster at J\"ulich Supercomputing Centre
(JSC), and on the GCS Supercomputers HAZEL HEN and HAWK
at H\"ochstleistungsrechenzentrum Stuttgart (HLRS).
The authors gratefully acknowledge the support of the Gauss Centre for
Supercomputing (GCS) and the John von Neumann-Institut für Computing
(NIC) for project CHMZ21,  CHMZ23 and HINTSPEC at JSC and project GCS-HQCD at HLRS.
This work has been supported by Deutsche Forschungsgemeinschaft
(German Research Foundation, DFG) through project HI 2048/1-2 (project
No.\ 399400745) and through the Cluster of Excellence ``Precision Physics,
Fundamental Interactions and Structure of Matter'' (PRISMA+ EXC
2118/1), funded within the German Excellence strategy (Project ID 39083149).
D.M.\ acknowledges funding by the
  Heisenberg Programme of the Deutsche Forschungsgemeinschaft (DFG, German
  Research Foundation) – project number 454605793.
A.G. received funding from the Excellence Initiative of Aix-Marseille
University - A*MIDEX, a French \emph{Investissements d'Avenir}
programme, AMX-18-ACE-005 and from the French National Research Agency
under the contract ANR-20-CE31-0016. We are grateful to our colleagues
in the CLS initiative for sharing ensembles.
Parts of the statistical data analysis have been performed using the $\Gamma$-method in the implementation of the \texttt{pyerrors} package \cite{Wolff:2003sm,Ramos:2018vgu,Joswig:2022qfe}.
}

{
\setlength{\bibsep}{.28em}
	
	\providecommand{\href}[2]{#2}\begingroup\raggedright\endgroup
	
}

\end{document}